\documentclass[12pt]{article}
\usepackage{amsmath, amssymb}
\usepackage{physics}
\usepackage{graphicx}
\usepackage{authblk}   % For blocks of authors

\def\fr{\frac}
\def\rarw{\rightarrow}

\def\nonu{\nonumber}

\def\om{\omega}
\def\gm{\gamma}

\def\lm{\lambda}
\def\eps{\epsilon}
\def\vareps{\varepsilon}
\def\al{\alpha}
\def\bt{\beta}

\def\dl{\delta}
\def\Dl{\Delta}
\def\sg{\sigma}

\def\Om{\Omega}

\def\rarw{\rightarrow}
\def\del{\partial}

\def\lan{\langle}
\def\ran{\rangle}
\def\beqr{ \begin{eqnarray}}
\def\eeqr{ \end{eqnarray}}
\def\beq{ \begin{equation}}
\def\eeq{ \end{equation}}
\def\bfig{\begin{figure}}
\def\efig{\end{figure}}

\begin{document}

\title{\mbox{} \hfill {\normalsize FERMILAB-PUB-18-391-APC} \\
  \mbox{} \hfill {\normalsize Accepted} \\
\mbox{}  \\
{\bf Micro-bunching for generating tunable narrow-band THz radiation at the FAST photoinjector}
}

% \href{https://urldefense.proofpoint.com/v2/url?u=http-3A__www.ctan.org_tex-2Darchive_macros_latex_contrib_elsarticle-257D-257BCTAN&d=DwICAg&c=gRgGjJ3BkIsb5y6s49QqsA&r=RUKcdRQc3Lj-y-vrxSo6vg&m=qAs9ZGSkRSY4i95e6L7_vmti4P2xOhWuJzYiSceqGKo&s=2K_aAdQeLDpf_2wvTlkdGT-Lzv9jmVreCVWd8iVsnmM&e=}.}

%% Group authors per affiliation:
\author[1]{J. Hyun}
\author[2, 3]{P. Piot}
\author[2]{T. Sen  \footnote{Corresponding author, email:  tsen@fnal.gov}}
\author[2]{J.C. Thangaraj}
  
\affil[1]{SOKENDAI, Tsukuba, Ibaraki 305-0801, Japan}
\affil[2]{Fermi National Accelerator Laboratory, Batavia, IL 60510, USA}
\affil[3]{Northern Illinois University, DeKalb, IL  60115, USA}
\date{}

\maketitle

\begin{abstract}
This paper presents expected THz radiation spectra emitted by micro-bunched electron beams produced using a slit-mask placed 
within a magnetic chicane in the FAST (Fermilab Accelerator Science and Technology) electron injector at Fermilab. Our purpose is 
to generate tunable narrow-band THz radiation with a simple scheme in a conventional photo-injector. Using the slit-mask in the
chicane, we create a longitudinally micro-bunched beam after the chicane by transversely slicing an energy chirped electron 
bunch at a location with horizontal 
dispersion. In this paper, we discuss the theory related to the micro-bunched beam structure, the beam optics,  the 
simulation results of the micro-bunched beam and the bunching factors. Energy radiated at THz frequencies from two sources:
coherent transition radiation and from a wiggler is calculated and compared. 
We also discuss the results of a  simple  method to 
observe the micro-bunching on a transverse screen monitor using a skew quadrupole placed in the chicane.
\end{abstract}

\section{Introduction} \label{Intro}
Accelerator based sources of THz radiation have been proposed and developed at several laboratories worldwide
\cite{Muller_2015, Hama_2011, 1-1,1-2,1-3,1-4,1-5,1-6,1-7,1-8,1-9,1-10,1-11,1-12}. 
THz radiation, whose frequency range is from 0.1 THz (wavelength $\lm=0.3$ mm) to 30 THz ($\lm=10\,\mu$m), is
non-ionizing and has high 
transmission through non-metallic materials such as clothes, paper, and plastic. Moreover, many materials have 
characteristic absorption spectra in this THz range. Therefore, THz radiation has been utilized in fundamental research 
in material, biological, and engineering science. For application to a wider range of fields such as industry, medicine and 
homeland security, a compact intense narrow-band THz source with tunable frequency is desired. 

Schemes to generate narrow-band THz radiation using laser-modulated electron beams have
been proposed \cite{Xiang_2009, Zhang_2017} which require special purpose modulator sections to modulate
the laser pulse acting on the electron beam. 
Here however, we focus on a simple method of generating a micro-bunched beam using standard components of a
photo-injector 
and a slit-mask \cite{1-13,1-14,Charles}. The transverse slicing of a bunch  by the mask is transformed into the longitudinal plane 
taking advantage of transverse dispersion at the slit-mask. We use a magnetic chicane consisting of four dipole 
magnets to either lengthen or shorten an electron bunch but in both cases create a beam with an appropriate comb structure 
required to generate THz radiation. 

We plan to perform the THz generation experiments at the Fermilab Accelerator Science and Technology (FAST) facility
 electron linac \cite{1-15,1-16}. The final goal is to produce a narrow-band THz wave with a frequency of over 1 THz and 
demonstrate that this method can provide a tunable narrow-band THz source. 
The frequency is tuned by choosing the RF phases in the cavities upstream of the chicane. Moreover, intense THz radiation 
can be  generated due to a high bunch repetition rate. 

In this paper, we present the theory related to the micro-bunched beam structure and simulation results of the micro-bunched 
beam, the expected THz spectra using coherent transition radiation (CTR) and a wiggler, as well as a method for observing the
 micro-bunched beam on a transverse screen monitor. In 
section \ref{FAST}, the FAST injector and the beam parameters are shown. In section \ref{Micro}, we 
describe the theory of the energy chirp, the width of the micro-bunches, the lowest frequency generated from the 
micro-bunched beam, and micro-bunch observation using a skew quadrupole magnet. The beam optics to generate the 
micro-bunched beam is shown in Section \ref{optics}. The expected micro-bunched beam structures, and the 
bunching factors are shown in Section \ref{sim}. A calculation of the THz radiation energy from CTR and a wiggler is 
presented in Section \ref{sec: radenergy}. 
The beam distributions after the chicane when a skew quadrupole is turned on for observing
the micro-bunching in the transverse plane are shown in Section \ref{obs}, and conclusions are presented in Section \ref{con}.

\section{FAST photoinjector }
\label{FAST}
\begin{figure}[t]
 \begin{center}
    \includegraphics[width=12.5cm, height=4.5cm]{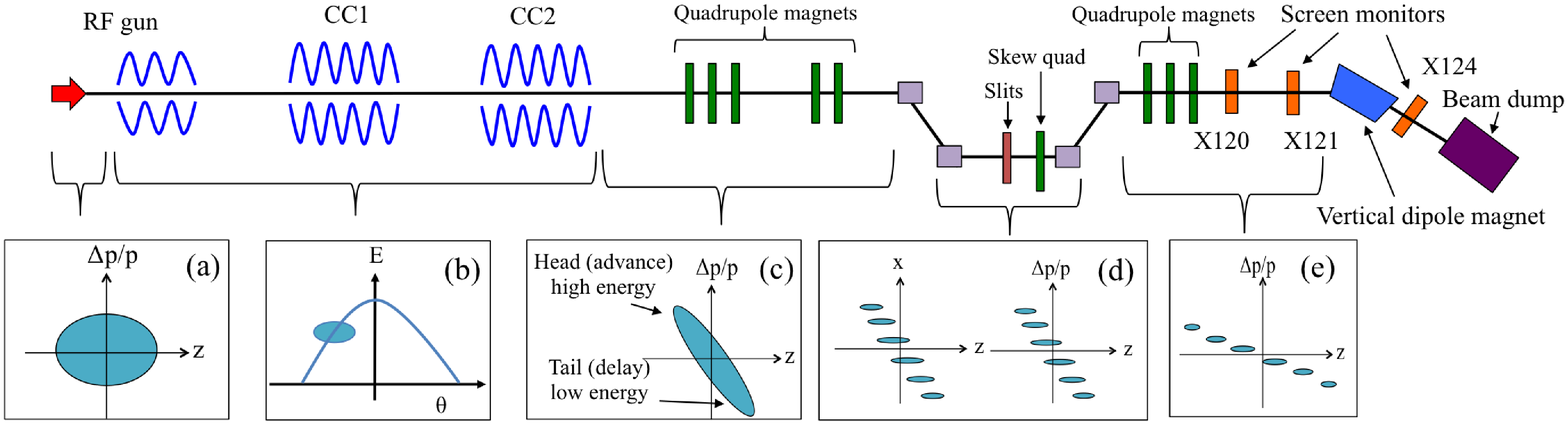}
    \caption{Layout of the FAST photoinjector.}
    \label{fig:1-1}
 \end{center}
\end{figure}

Figure \ref{fig:1-1} shows the layout of the FAST photoinjector. The injector consists of
an RF gun with a normal conducting 1.5 cell cavity, two TESLA style 9-cell superconducting cavities (CC1 and CC2), a 
magnetic chicane, a vertical dipole magnet for beam extraction, and a beam dump. A molybdenum disk coated with $\mathrm{Cs_2Te}$ is used as the cathode, and the RF gun is similar to the one developed for the FLASH facility at 
DESY \cite{2-1}. Electron bunches are emitted with a repetition rate of 3 MHz within a macro-pulse that lasts 
1 ms. The RF gun and the two superconducting cavities operate at an RF frequency of 1.3 GHz with a repetition rate of 
5 Hz. The main machine and beam parameters are shown in Table\,\ref{table: parameters}.
\begin{table}[h]
\caption{Relevant beam and machine parameters }
\begin{center}
\begin{tabular}{cc} \hline
 Parameter &  Value  \\ \hline\hline
Beam energy after gun &  5 MeV \\
Normalized emittance & $\sim$ 2 mm-mrad \\
Nominal bunch charge & 200 pC \\
rms bunch length & 0.9 mm \\
Uncorrelated rms energy spread & 0.1 \% \\
Peak operational gradients in (CC1, CC2)  &  (16 , 20) MV/m \\
Fixed beam energy after CC2 & $\sim$35 MeV \\
Operational range of rf phases &  $-35^\circ \leftrightarrow 35^\circ$ \\
Slits (spacing $D$, width $W$, thickness $t$) & (0.95, 0.050, 0.50) mm \\
Chicane dipole bend radius $\rho$, angle $\theta$ &  0.84 m, 18$^{\circ}$  \\
Chicane longitudinal dispersion $RC_{56}$ & -0.18 m \\
Chicane horizontal dispersion $\eta$ & -0.34 m \\
\hline
\end{tabular}
\label{table: parameters}
\end{center}
\end{table}
The electron beam sizes are controlled by a doublet and two triplets of quadrupoles installed in the 
beamline. The electron beam sizes are measured using two YAG screen monitors (at X120, X121 in Figure \ref{fig:1-1}) downstream of the 
chicane. When the micro-bunched beam hits an Al foil at X121, THz radiation is emitted as coherent transition radiation 
(CTR), and it can be measured using a pyrometer or a bolometer \cite{1-9,1-17}.  

\section{Micro-bunched beam production}
\label{Micro}
In this section, we present the theory and the method of creating and observing a micro-bunched beam using an energy chirped bunch, a slit-mask and a skew quadrupole 
in the magnetic chicane.

\subsection{Energy-chirped beam}
An energy chip represents a correlation between longitudinal position $z$ and energy deviation $\delta$, and is defined by 
$h=\partial\delta/\partial z|_{z=0}$. This correlation can be
 produced by accelerating the beam with an off-crest rf phase in a cavity (see (a), (b), and (c) in Fig.\,\ref{fig:1-1}). 

We define $E_{0, ref}, E_{1, ref}, E_{2, ref}$ to be the nominal beam energies at the entry of CC1, exit of CC1, and the exit of 
CC2 respectively. The reference particle
at the longitudinal center ($z=0$) of the bunch has the nominal energy while a particle at an arbitrary position $z$ has energies $E_1(z), E_2(z)$ after CC1 and CC2 respectively.
      \begin{eqnarray}
E_1(z) & = &  E_{0, ref}(1 + \dl_0) + e  V_{1, rf} \cos (\phi_{1, rf} - k_{ rf} z), \\
E_2(z) & = &  E_{1}(z) + e  V_{2, rf} \cos (\phi_{2, rf} - k_{ rf} z), 
     \label{eq:3-1-1}
    \end{eqnarray}
where $\dl_0$ is the initial energy deviation (due to the uncorrelated energy spread from the rf gun), $(V_{1,rf}, V_{2, rf})$ 
are the the voltages  and
$(\phi_{1, rf}, \phi_{2, rf})$ are the rf phases in CC1, CC2 respectively. $k_{rf}$ is the rf wave number, the same for both cavities. The relative energy
deviations from the nominal energy after CC1 and CC2 are given by $\dl_1(z) = (E_1(z)- E_{1,ref})/E_{1,ref}$ and $\dl_2(z) = (E_2(z)- E_{2,ref})/E_{2,ref}$
respectively. Therefore, the energy chirp $h_1$ after CC1  and the total chirp $h_T$ after CC2 can be written as
      \begin{eqnarray}
h_1 &  \equiv & \frac{\del \dl_1}{\del z}|_{z=0} = \frac{k_{rf}  v_1 \sin \phi_{1, rf}}{1 +  v_1 \cos \phi_{1, rf}},  \label{eq: h1} \\
h_T &  \equiv & \frac{\del \dl_2}{\del z}|_{z=0} = \frac{k_{rf}[ v_1 \sin \phi_{1, rf} + v_2 \sin \phi_{2, rf}]}{1 +  v_1 \cos \phi_{1, rf} + 
  v_2 \cos \phi_{2, rf}},   \label{eq: hT} 
      \end{eqnarray}
where $v_i = e V_{i, rf}/E_{0, ref}, i = 1, 2$ are dimensionless parameters. 
In Eqs.\,(\ref{eq: h1}, \ref{eq: hT}). $h<0 $(respectively $h>0$) means that higher energy (respectively lower energy) electrons are in 
the bunch head  after the cavity, which leads to bunch lengthening (respectively shortening) 
(see (c) in Fig.\,\ref{fig:1-1}).  The energy chirp for the maximum compression of an electron beam is 
$h=-1/RC_{56} \simeq 5.6\,\mathrm{m^{-1}}$ at FAST, where $RC_{56}$ is the longitudinal dispersion generated by the chicane. 

The energy chirp can be produced by accelerating with an off-crest phase in either one or both cavities. 
Since off-crest acceleration lowers the beam energy, the 
two accelerating voltages are changed to keep the final energy fixed at 35 MeV which is the value for 
$\phi_{1, rf}=\phi_{2, rf}=\pm 35^\circ$ at the peak voltage gradients shown in Table \ref{table: parameters}. The constant 
beam energy simplifies operation by requiring no tuning of the dipole strengths in the chicane for each choice of rf phases. 
The energy chirps calculated with Eqs.\,(\ref{eq: h1}, \ref{eq: hT}) are summarized in Table\,\ref{tb:1} in Section\,\ref{sim}.
 
\subsection{Micro-bunched beam}
The energy-chirped beam from the two cavities is sent to the  chicane where the horizontal dispersion has negative
values. Electrons in the chicane are separated horizontally with higher energy (lower energy) electrons passing through the 
inside (outside) of the ideal orbit. The slit-mask in the middle of the chicane splits the beam horizontally into sections 
(see (d) in Fig.\,\ref{fig:1-1}). Beam transmission through the slit-mask is expected to be around 5\% from the ratio of
the slit width $W$ to the slit spacing $D$. The particles passing through a slit opening are fully transmitted  since the
beam divergence at the slit-mask (= 0.7 mrad) is much smaller than the opening angle  $W/t = 0.1$ rad where $t$ is the
thickness of the slit, while the particles passing
through the tungsten are scattered at large angles and lost downstream of the mask. This transmission ratio has been 
confirmed with {\sc Geant4} simulations \cite{geant}. 

In the bunch lengthening mode of operation, higher energy electrons after the chicane are at
the bunch head while the lower energy electrons are at the bunch tail. The horizontally separated bunch after the slit-mask
is transformed into a longitudinally separated beam (or into  micro-bunches) after the chicane (see (e) in 
Fig.\,\ref{fig:1-1}). The lengthening increases the longitudinal separation between the micro-bunches. 

The longitudinal width and spacing of the micro-bunches can be found using the transfer matrix. The 6$\times$6 transfer matrix of the 
dogleg (from the entrance of the chicane to the slit-mask in the middle of the chicane) composed of rectangular bend magnets 
transforms the six dimensional phase space variables $(x, x', y, y', z, \dl)$ as 
\begin{eqnarray}
\left(
\begin{array}{c}
x \\
x' \\
y \\
y' \\
z  \\
\delta
\end{array}
\right)_s
=
\left(
\begin{array}{cccccc}
1 & R_{12} & 0 & 0 & 0 & R_{16} \\
0 & 1 & 0 & 0 & 0 & 0            \\
0 & 0 & R_{33} & R_{34} & 0 & 0            \\
0 & 0 & R_{43} & R_{44} & 0 & 0                       \\
0 & R_{52} & 0 & 0 & 1 & R_{56}  \\
0 & 0 & 0 & 0 & 0 & 1  \\
  \end{array}
 \right)   
 \left(
\begin{array}{c}
x \\
x' \\
y \\
y' \\
z  \\
\delta
\end{array}
\right)_1.   
\end{eqnarray}
The matrix elements are given by 
\begin{eqnarray}
R_{12} & = & d_1 + \frac{1}{2} d_2+2\rho\sin\theta, \;\;\; R_{16} = - 2(d_1+\rho\sin\theta)\tan(\frac{\theta}{2})  \nonu \\
R_{33} & = & \cos2\theta-\frac{(d_1+d_2)}{2\rho}\sin2\theta+\frac{d_1d_2}{2\rho^2}\sin^2\theta  \nonu \\ 
R_{34} & = & d_1\cos^2\theta+\rho\sin 2\theta+\frac{d_2}{2}\cos2\theta-\frac{d_1 d_2}{4\rho}\sin2\theta  \nonu \\  
 R_{43} & = & \fr{d_1}{\rho^2}\sin^2\theta  - \fr{1}{\rho}\sin2\theta,  \;\;\; R_{44} =\cos2\theta-\fr{d_1}{2\rho}\sin2\theta, \;\;\; R_{52}=R_{16},  \nonu \\
 R_{56} & =  & 4(d_1 + 2 \rho \sin\theta)\tan^2\fr{\theta}{2} -2\rho(\bt_k^2 \theta-\sin\theta)+\frac{1}{\gamma^2}(d_1 + \frac{1}{2} d_2).
\label{eq:Rmat_dog}
\end{eqnarray}
In the above, we assumed a symmetric chicane with the same magnitude of the bend angle $\theta$ and bend radius $\rho$
in each dipole and where the separation $d_1 $  between the first and second dipoles is the same as
that between the third and fourth dipoles and $d_2$ is the separation between the second and third dipoles. 
$\gamma$ is the beam's relativistic Lorentz factor in the chicane. This is large enough that the third term in
$R_{56}$ is negligible. We find that second order chromatic terms such as $T_{166}, T_{566}$ are larger than but are of the same 
order of magnitude as the first order terms, so their influence on the dynamics is likely to be negligible for the usual beam
energy spreads. 

The horizontal position of a particle at the slit-mask is given by $x_s=x_c+R_{16}\delta_1$, where 
$x_c=x_1+R_{12}x'_1$, and where $(x_1, x'_1)$ are the horizontal coordinates of the particle at the chicane's entrance.
The energy deviation (constant through the chicane) is $\delta_1 (=\delta_s)= \delta_0+ h z_1$, where $\delta_0$ is the 
deviation due to the uncorrelated energy spread, $h$ is the energy chirp, and $z_1$ is the longitudinal coordinate of the particle at the chicane entrance.
In terms of the transverse positions and $\dl_0$, this can be written as
 \begin{eqnarray}
     \label{eq:3-2-2}
 z_1=\frac{(x_s-x_c)}{h R_{16}}-\frac{\delta_0}{h}.
 \end{eqnarray}

The transfer matrix $RC$ for the complete chicane can be similarly written down. For our purposes, the 
non-zero elements of interest are in the $(x,x',z,\dl)$ planes
       \begin{eqnarray}
RC_{11} = 1, \;\;\; RC_{12} = 2 R_{12}, \;\;\;  RC_{22} = 1,  \nonu \\
RC_{55} = 1, \;\;\; RC_{56} = 2 R_{56}, \;\;\;  RC_{66} = 1 \; ,
       \end{eqnarray}
where $R_{12}, R_{56}$ are given by the expressions in Eq.\,(\ref{eq:Rmat_dog}). 
The other non-zero elements $(RC_{33}, RC_{34}, RC_{43}, RC_{44})$ are not needed here and are omitted. 

The longitudinal coordinate $z_2$ after the chicane is 
       \begin{eqnarray}
       \label{eq:3-2-3}
  z_2 & = & z_1+RC_{56}\delta_1  =\frac{(1+ h RC_{56})}{h R_{16}}(x_s-x_c)   - \frac{\delta_0}{h} 
\nonu \\
& = & \frac{(1+ h RC_{56})}{h R_{16}}(x_s-x_1) - \frac{ R_{12}(1+ h RC_{56})}{h R_{16}}x_1' - \fr{\dl_0}{h}.
       \end{eqnarray}
In Eq.(\ref{eq:3-2-3}), only the variables $x_1, x_1'$ have a non-zero covariance, the other variables are uncorrelated. 
Since the full slit width $W$ is smaller than the the rms  beam size, we can assume that the beam distribution in 
the slits is uniform. We have for the variances and the non-zero covariance, 
\begin{eqnarray}
\lan x_s^2 \ran & = & \fr{W^2}{12}, \;\;\; \;\;  \lan \delta_0^2  \ran  =   \sg^2_{\dl, U}  \\ 
\lan x_1^2 \ran &  = &  \bt_1 \vareps, \;\;\; \lan (x_1')^2 \ran =   \fr{1 + \al_1^2}{\bt_1}\vareps, \;\;\; 
\lan x_1 x_1' \ran = - \al_1 \vareps 
\end{eqnarray}
where $\bt_1, \al_1$ are the horizontal beta and alpha functions at the chicane entrance, $\vareps$ is the  
un-normalized horizontal emittance, and $\sg_{\dl, U}$ is the 
uncorrelated rms relative energy spread in the chicane. We obtain the  rms length of a micro-bunch $\sigma_{z_2, MB}$ from
 \begin{eqnarray}
\sg_{z, MB}^2 & = & \left[\frac{(1+ h RC_{56})}{h\eta}\right]^2   \times 
\left\{ \fr{W^2}{12} +  \vareps \left[\bt_1 -  2 R_{12}  \al_1 + R_{12}^2 \fr{1 + \al_1^2}{\bt_1}\right] \right\}  + 
 \fr{\sigma^2_{\dl, U}}{h^2}   \nonu \\
& = & \left[\frac{(1+ h RC_{56})}{h\eta}\right]^2   \times 
\left\{ \fr{W^2}{12} +  \vareps \bt_S\right\}  +  \fr{\sigma^2_{\dl, U}}{h^2},  \label{eq:sigma_MB}
 \end{eqnarray}
where $\eta=R_{16}= - 2 (d_1 + \rho \sin\theta) \tan(\theta/2)$ is the dispersion at the slits and 
$\bt_S= \bt_1 -  2 R_{12}  \al_1 + R_{12}^2 (1 + \al_1^2)/\bt_1 $ is the beta function at the slits. 
In most cases, the contribution of the betatron size dominates, so that we have approximately 
 \begin{eqnarray}
\sg_{z, MB} \approx \fr{|(1 +  h RC_{56})|}{|\eta h|} \sqrt{\bt_S \varepsilon}  \; .
 \end{eqnarray}
Eq.\,(\ref{eq:sigma_MB}) shows that $\varepsilon\beta_S$ and $\sigma_{\dl, U}$ should be small to minimize the length of each
micro-bunch and therefore  create larger longitudinal separations between the micro-bunches. 

Denoting the horizontal position at the $i$th slit by $x_s^i$, we have on average $\lan x_s^i - x_s^{i-1}\ran= D$ where 
$D$ is the slits spacing, while $\lan x_c^I - x_c^{I-1} \ran = 0$.
Hence the average longitudinal separation $\lan \Delta z \ran $ after the chicane   between particles which pass through 
neighboring slits is
\begin{eqnarray}
       \label{eq:3-2-4}
|\lan   \Delta z \ran | = |\lan( z^i_2-z^{i-1}_2 )\ran|=\frac{D}{|h\eta|}(|1+ h RC_{56}|). 
   \end{eqnarray}
Here we have dropped the  negligible differences in energy between particles at neighboring slits. 
The micro-bunched beam's widths and spacings computed for each energy chip are summarized in Table\,\ref{tb:1} in 
Section\,\ref{sim}.

\subsection{Frequency dependence on energy chirp}
\begin{figure}[t]
 \begin{center}
    \includegraphics[width=8cm, height=5cm]{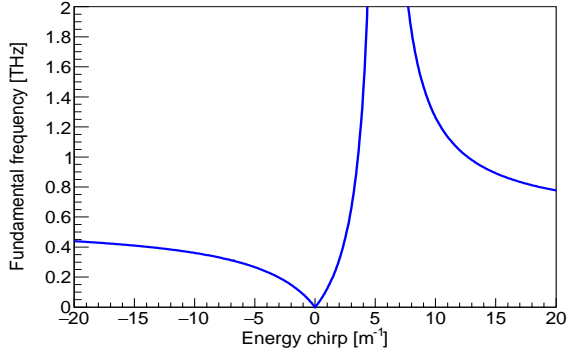}
    \caption{Fundamental frequency depending on energy chirps. The fundamental frequency becomes large near the
 maximum compression at $h=5.6$ m$^{-1}$. }
    \label{fig:2}
 \end{center}
\end{figure}

THz radiation can be generated by allowing the micro-bunched beam to traverse an Al foil.  The fundamental frequency 
$f_0$ is determined by the separation between the micro-bunches for a comb structure beam and is given by
\begin{eqnarray}
       \label{eq:3-3-1}
      f_0=\frac{c}{|\lan \Delta z \ran| }=\frac{ c|h\eta|}{D|1+ h RC_{56}|}.
 \end{eqnarray}
This equation is valid as long as the separation satisfies $|\lan \Dl z \ran| \gg RC_{56} \sg_{\dl, U}$ which is generally true, except
in the vicinity of maximum compression where $\lan \Dl z \ran \rarw 0$. 
Figure\,\ref{fig:2} shows the fundamental frequency as a function of the energy chirp. The fundamental frequency can be 
changed by varying the  energy chirp.  From Fig.\,\ref{fig:2}, the fundamental frequencies are about 0.3 THz, 0.33 THz, and 0.38 THz at 
negative chirps $h$=-7, -9, and -16 $\mathrm m^{-1}$, respectively, and 1.82 THz, 1.27 THz, and 0.77 THz at positive chirps 
$h$=7, 9, and 
16 $\mathrm m^{-1}$, respectively. The fundamental frequency is zero when there is no chirp (at $h=0\,\mathrm m^{-1}$ )
and goes to large values close to maximum compression as $h\rarw -1/(RC_{56}) \sim 5.6\,\mathrm m^{-1}$. While positive $h$ values 
lead to larger fundamental frequencies, they also compress the entire bunch structure and lead to overlap between micro-bunches, 
and the frequency spectra are broadband rather than narrow-band. 
Figure\,\ref{fig:2} also shows that the fundamental frequency changes slowly beyond $|h| \approx 20$ m$^{-1}$, so there
is no advantage in going beyond these chirp values. The fundamental frequencies
computed for each energy chirp are summarized  in Table\,\ref{tb:1} in Section \ref{sim}.

\subsection{Observing micro-bunching in the transverse plane}
A skew quadrupole magnet installed in the chicane can be used to confirm that a micro-bunched beam is produced
prior to detecting the radiation. When the 
skew quadrupole is turned on in the chicane where the horizontal dispersion is non-zero, vertical dispersion is generated 
downstream of the 
skew quadrupole via beam coupling. Due to the vertical dispersion after the chicane, the information on the beam separation in 
the horizontal plane (energy-plane) at the skew quadrupole is transferred to the vertical plane \cite{3-1}. Using a 
Yttrium Aluminum Garnet (YAG) scintillating  screen 
downstream of the chicane, we can observe the electron beam separation in the vertical plane. The vertical spacing can be 
found through the transfer matrix from the skew quadrupole in the chicane to the monitor downstream of the chicane. The 
phase space vectors at the monitor and slit locations are related via
\begin{eqnarray}
\left(
\begin{array}{c}
x \\
x' \\
y \\
y' \\
z  \\
\delta
\end{array}
\right)_M
& = & RM \cdot
\left(
\begin{array}{cccccc}
1 & 0 & 0 & 0 & 0 & 0   \\
0 & 1 & -k_S & 0 & 0 & 0   \\
0 & 0 & 1 & 0 & 0 & 0   \\
-k_S & 0 & 0 & 1 & 0 & 0   \\
0 & 0 & 0 & 0 & 1 & 0   \\
0 & 0 & 0 & 0 & 0 & 1   \\
  \end{array}
 \right)
 \left(
\begin{array}{c}
x \\
x' \\
y \\
y' \\
z  \\
\delta
\end{array}
\right)_s.  
\end{eqnarray}
The non-zero components of the matrix $RM$ are
\begin{eqnarray}
RM_{11} & = & 1, \;\;\; RM_{12} = R_{12}+ d_M, \;\;\; RM_{16} =  - R_{16}, \;\;\; RM_{22} = 1,  \nonu \\
RM_{33} & = & \cos 2\theta - \fr{d_1}{2\rho}\sin 2\theta + \fr{d_M}{\rho^2}(d_1 \sin^2\theta - \rho \sin 2\theta )  \nonu  \\
RM_{34} & = &  R_{34} +  d_M R_{33} \nonu \\
RM_{43} & = &  R_{43}   , \;\;\;  RM_{44} =\cos2\theta + \fr{1}{2\rho^2}[ d_1 d_2 \sin^2 \theta - (d_1 + d_2)\rho \sin 2\theta]
\nonu  \\
RM_{52} & = & -R_{52}, \;\;\; RM_{55} = 1, \;\;\; RM_{56} =    R_{56} + \fr{d_M}{\gm^2}, \;\;\;  RM_{66} = 1,
\label{eq: mat_mon}
\end{eqnarray}
where $k_S$ is the inverse focal length of the skew quadrupole, $d_M$ is the distance from the end of the chicane to the 
monitor, and the matrix elements
$R_{ij}$ are those in Eq.\,(\ref{eq:Rmat_dog}). The vertical position $y_M$ at the monitor after the chicane is,
       \begin{eqnarray}
       \label{eq:3-4-2}
         y_M & = & -k_S  RM_{34} x_s+RM_{33}y_s+RM_{34}y'_s=y_{2, k_S=0} - k_S RM_{34} x_s.\\
         y_{2, k_S=0} & = & RM_{33} y_s+RM_{34}y'_s.  \nonu 
        \end{eqnarray} 
Taking the effect of the slit-mask into account, the average vertical spacing is
  \begin{eqnarray}
 \lan \Delta y \ran & = &  \lan (y^i_{M_{k_S=0}}-y^{i-1}_{M_{k_S=0}}) - k_S RM_{34}(x^i_s-x^{i-1}_s) \ran  \nonu \\
  \Rightarrow | \lan \Delta y \ran |  & = & | k_S| D \; RM_{34} \; .         \label{eq:3-4-3}
  \end{eqnarray}
where $D$ is the horizontal spacing of the slits and we used $\lan (y^i_{M_, k_S=0} - y^{i-1}_{M_, k_S=0}) \ran = 0$. The vertical 
average spacing is proportional to the strength of the skew quadrupole, increases with the distance $d_M$ to the monitor but
is independent of the chirp.  The electron beam should be focused vertically at the monitor to observe clearly separated 
slit images because the separation, determined by the second term in Eq.\,(\ref{eq:3-4-2}), should be larger 
than the first term of this equation which is determined by the betatron beam size.

The slope of the vertical separation with skew quadrupole strength can be used to infer the longitudinal separation that would be produced in the absence of this quadrupole via
  \begin{eqnarray}
   |\lan   \Delta z \ran| = \left[\fr{|1 +  h RC_{56}|}{|h \eta|}\right]  \left( \fr{|\lan \Dl y\ran |}{|k_S| RM_{34}} \right), 
   \end{eqnarray}
where the terms in square brackets depend on the energy chirp and the chicane while those in parentheses depend on
the observations at the transverse screen monitor.

\section{Beam optics for the micro-bunched beam}
\label{optics}

\begin{figure}[t]
 \begin{center}
 \includegraphics[width=10cm, height=6cm]{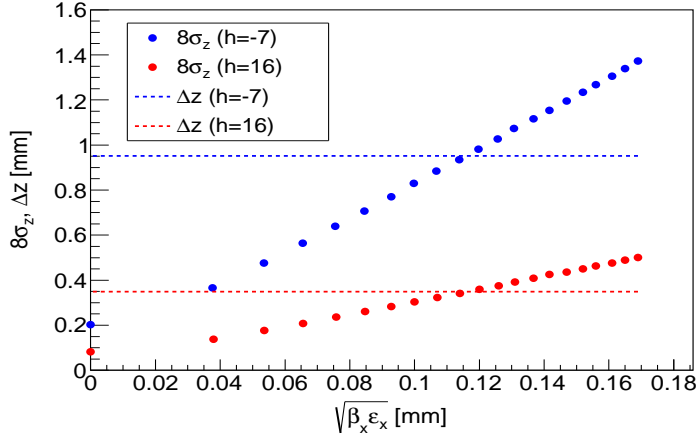}
    \caption{Spacing ($8\sigma_z$) and width ($\Delta z$) of micro-bunched beams as a function of 
$\sqrt{\beta_x\varepsilon_x}$ at the slit-mask for different energy chirps. Dots and lines represent width and spacing of micro-bunched beams, respectively. }
    \label{fig:3}
 \end{center}
\end{figure}

In this section, we show the beam optics to produce a micro-bunched beam. The beam optics and particle tracking from the 
entrance of CC1 to the beam dump were simulated using  {\sc elegant} \cite{4-1}. As initial parameters before CC1, we used 
the beam parameters shown in Table \ref{table: parameters} while the initial Twiss functions were 
$\beta_x=\beta_y= 4.89\,\mathrm{m}$, and $\alpha_x=\alpha_y=0$. 
These beta functions are chosen so that the electron beam sizes at the entrance of CC1 are 1 mm in each plane. The
accelerating voltages in CC1 and CC2 are tuned so that the electron beam energy stays constant at  35 MeV, as mentioned in
Section\,\ref{Micro}.

Clear separations of the micro-bunches after the chicane require that the total width of a micro-bunch  to be smaller 
than the micro-bunch spacing $\lan \Delta z \ran$. We set $8\sigma_{z, MB}\sim \lan \Delta z \ran$ by controlling $\beta_x$ at 
the slit-mask. 
Figure\,\ref{fig:3} shows the relation between total width ($8\sigma_z$) of micro-bunched beams and betatron beam 
size $\sqrt{\varepsilon_x\beta_{x,S}}$ for different energy chirps. Dots and lines represent widths and spacings of micro-bunched 
beams, respectively. From Fig.\,\ref{fig:3}, we chose $\beta_{x,S}=0.5\,\mathrm m$ at the slits so that  the betatron beam 
size $\sqrt{\varepsilon_x\beta_{x,S}}=0.12 $ mm at the intersection of $8\sigma_z$ and $\lan \Delta z \ran$ for energy 
chirps except 
for those close to $h=+5.6\,\mathrm m^{-1}$ at the maximum compression where $\lan \Delta z \ran =0$. When the energy 
chirps are 
$h=+7, +9\,\mathrm m^{-1}$,  $\lan \Delta z \ran$ is still quite small.  Therefore, correspondingly small values of 
$\varepsilon_x\beta_x$ and uncorrelated energy spread $\sg_{\dl, U}$ are required to obtain a clearly separated longitudinal 
distribution after the chicane.

\begin{figure}[t]
 \begin{center}
    \includegraphics[width=8cm, height=6cm]{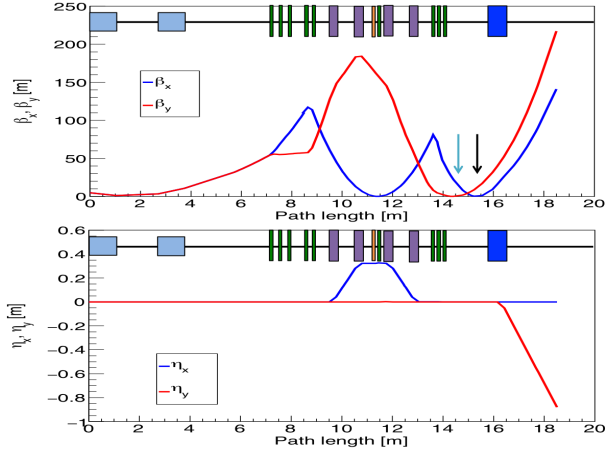}
    \caption{Beta and dispersion functions at $h=-7\,\mathrm{m^{-1}}$ for CC1 chirp. Blue and red lines show horizontal and vertical planes respectively. Light blue, green, magenta boxes show the accelerating structures, (normal and skew) quadrupole magnets, and dipole magnet of the chicane, respectively. The navy blue box at $\sim 16$m represents the vertical bend magnet which sends
the beam to the dump.}
 \end{center}
    \label{fig:4}
\end{figure}

Figure\,\ref{fig:4} shows the beam optics from CC1 to the beam dump at $h= -7\,\mathrm{m^{-1}}$  with a  chirp only in CC1. 
The beam optics for different energy chirps shows similar behavior. The horizontal beta function for all cases is focused to  
about 0.5 m at the slit-mask. We also focused the vertical beam size at the screen monitor X120 downstream of the chicane 
to be as small as possible for clear separations of the vertical slit images when the skew quadrupole is turned on. 

\section{Simulations of micro-bunched beams}
\label{sim}

\begin{figure}[t]
 \begin{center}
  \begin{tabular}{c}
   \begin{minipage}{0.5\hsize}
    \begin{center}  
\includegraphics[width=6.cm, height=5.0cm]{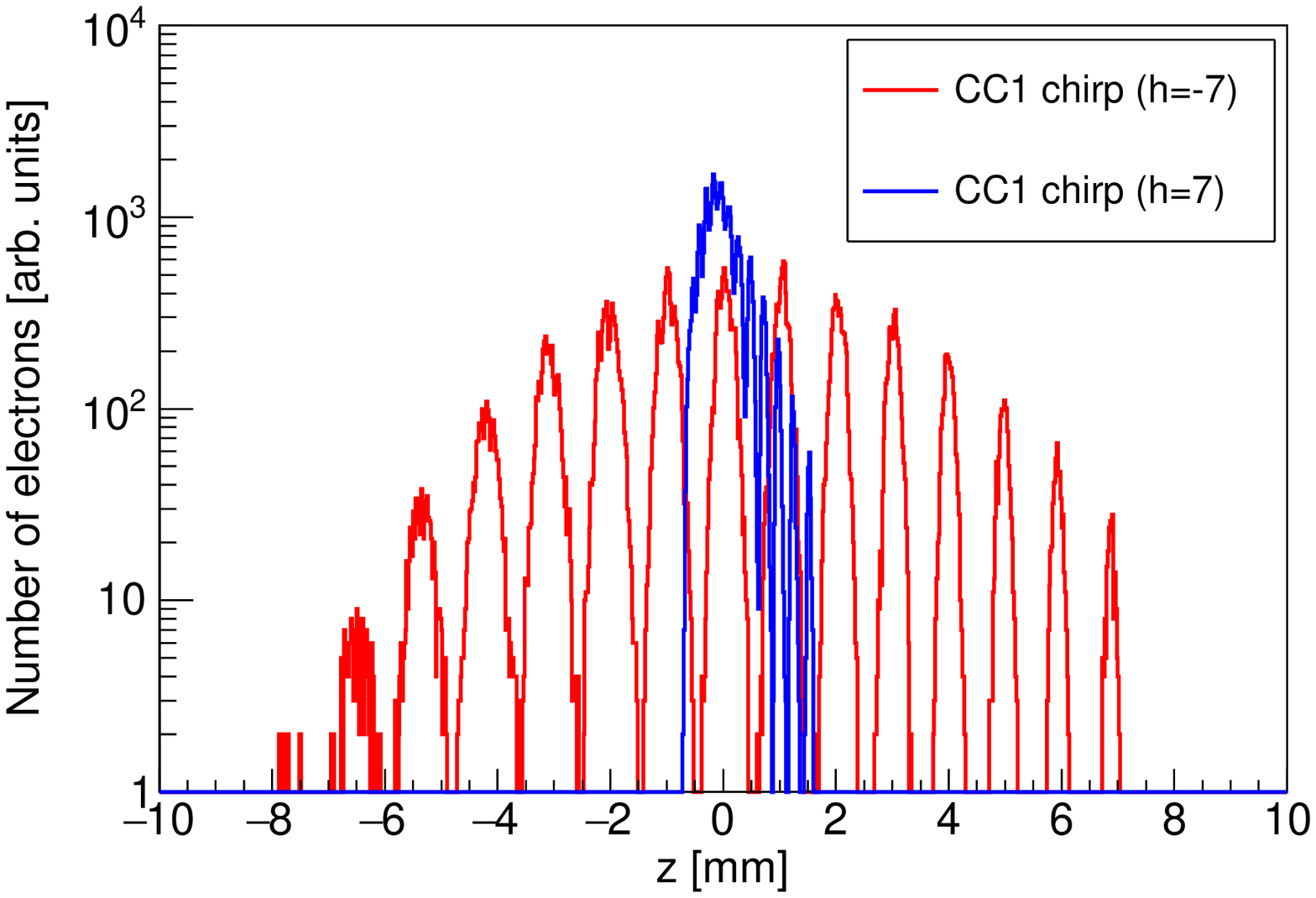}   
 \end{center}
\end{minipage}
\begin{minipage}{0.5\hsize}
 \begin{center}
\includegraphics[width=6.cm, height=5.0cm]{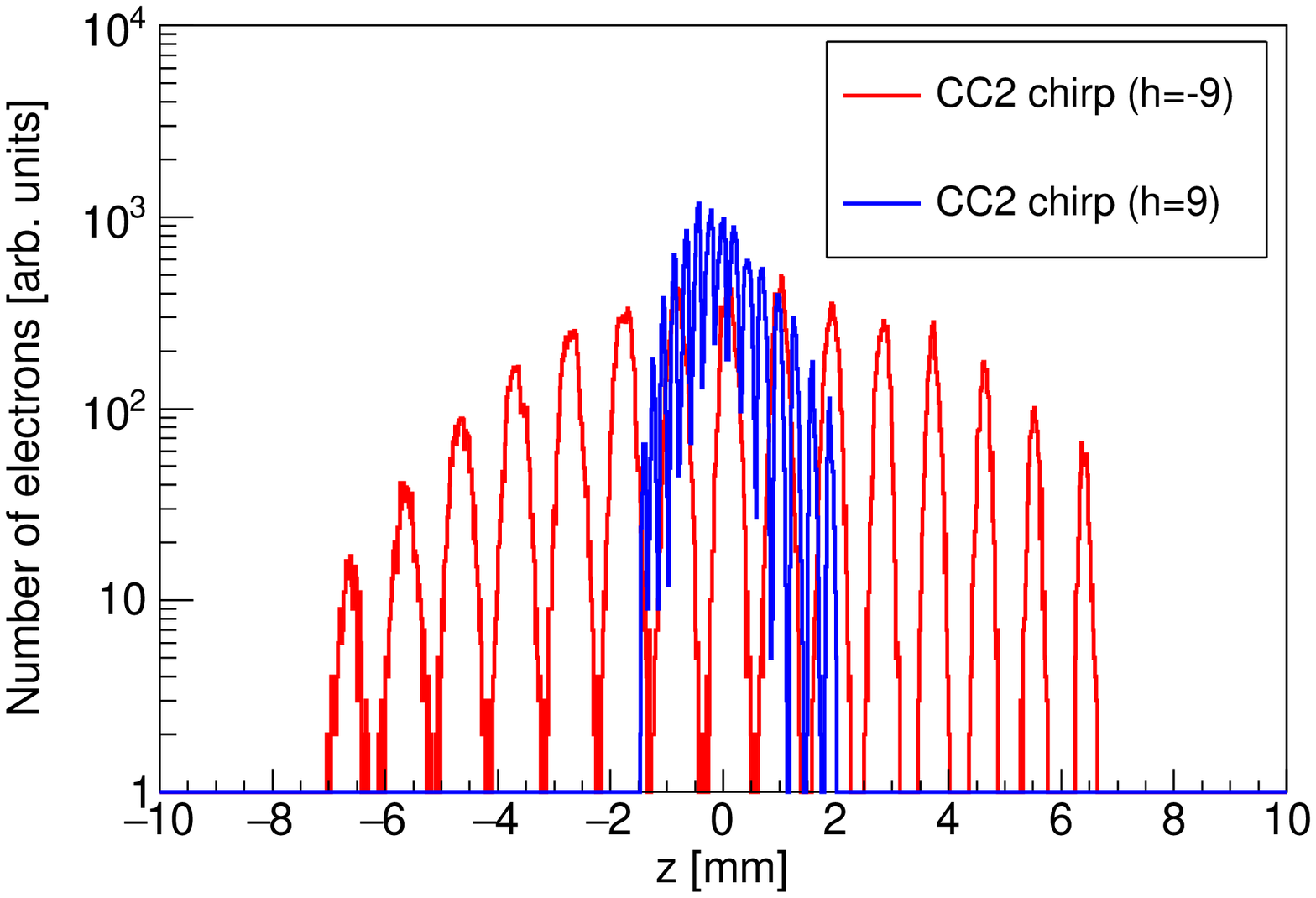}   
 \end{center}
\end{minipage}
\end{tabular}
\centering
  \includegraphics[width=6.cm, height=5.0cm]{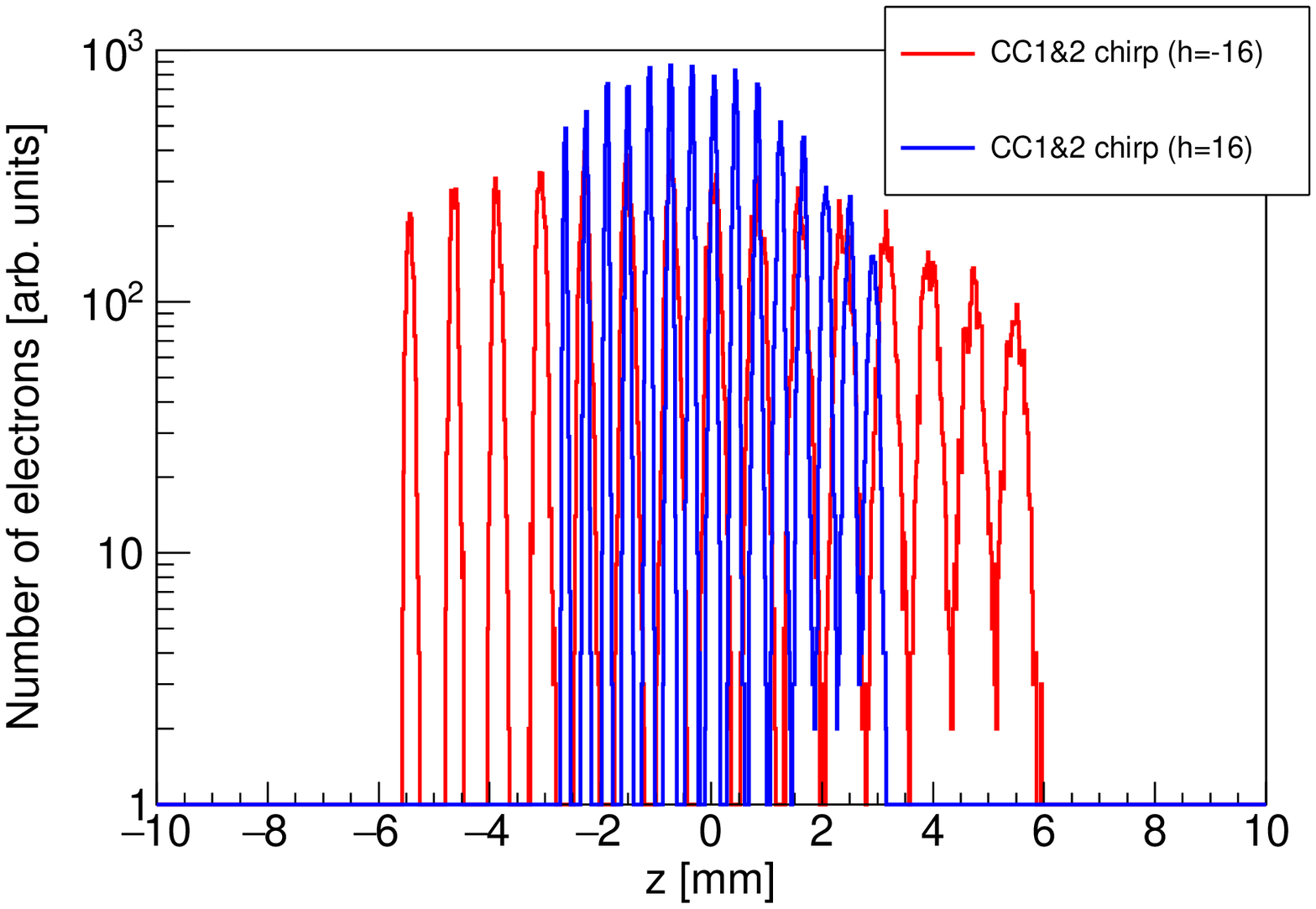}
\caption{Longitudinal distributions at CC1 chip ($h=\pm7\,\mathrm{m^{-1}}$), CC2 chirp ($h=\pm9\,\mathrm{m^{-1}}$), and CC1\&CC2 chips ($h=\pm16\,\mathrm{m^{-1}}$). Red 
lines are the bunch lengthening mode, and blue lines are the over compressed mode.}
\label{fig:5-1}
\end{center}
\end{figure}

\begin{figure}[t]
 \begin{center}
  \begin{tabular}{c}
   \begin{minipage}{0.5\hsize}
    \begin{center}  
      \includegraphics[width=6.0cm, height=5.0cm]{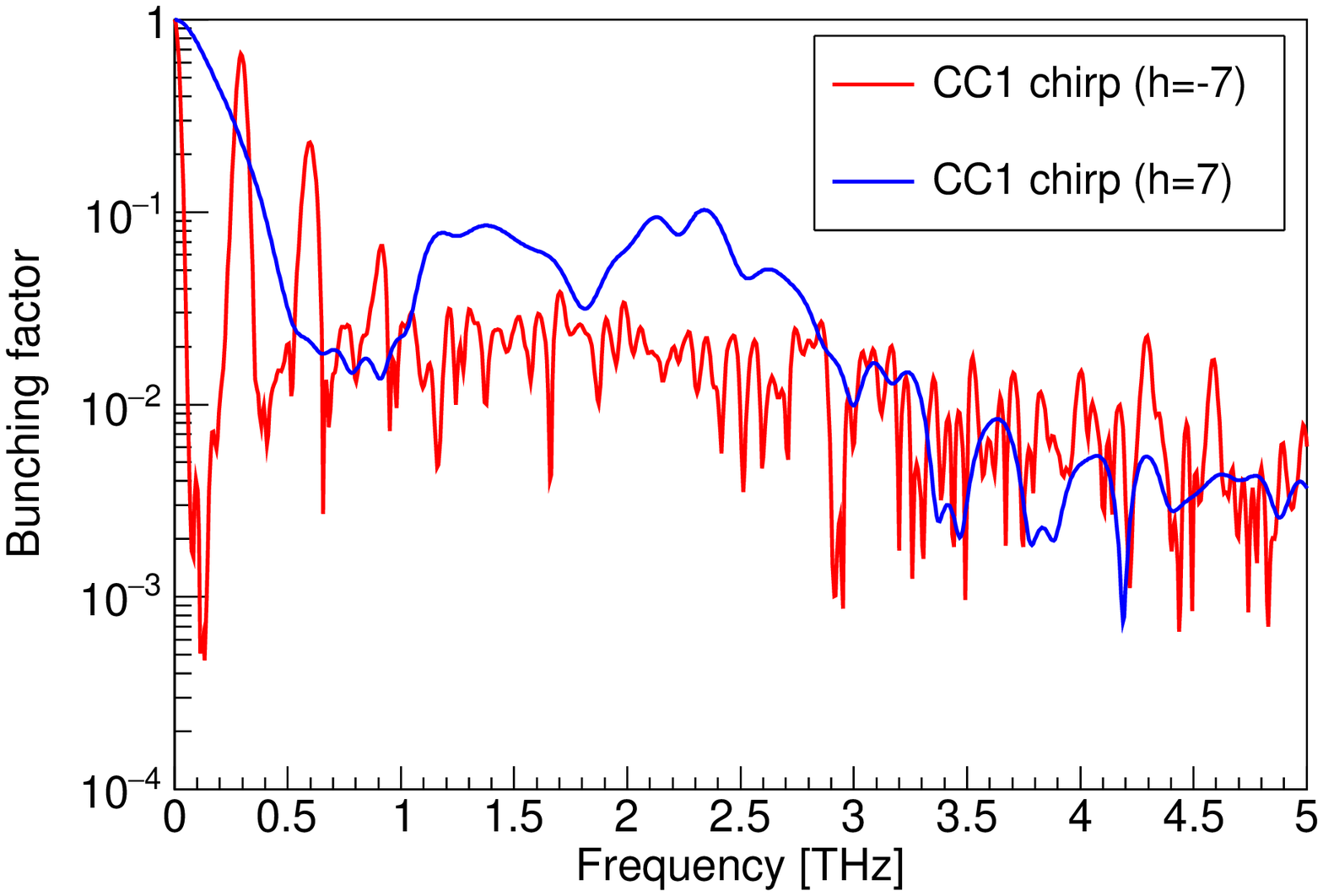}   
 \end{center}
\end{minipage}
\begin{minipage}{0.5\hsize}
 \begin{center}
      \includegraphics[width=6.0cm, height=5.0cm]{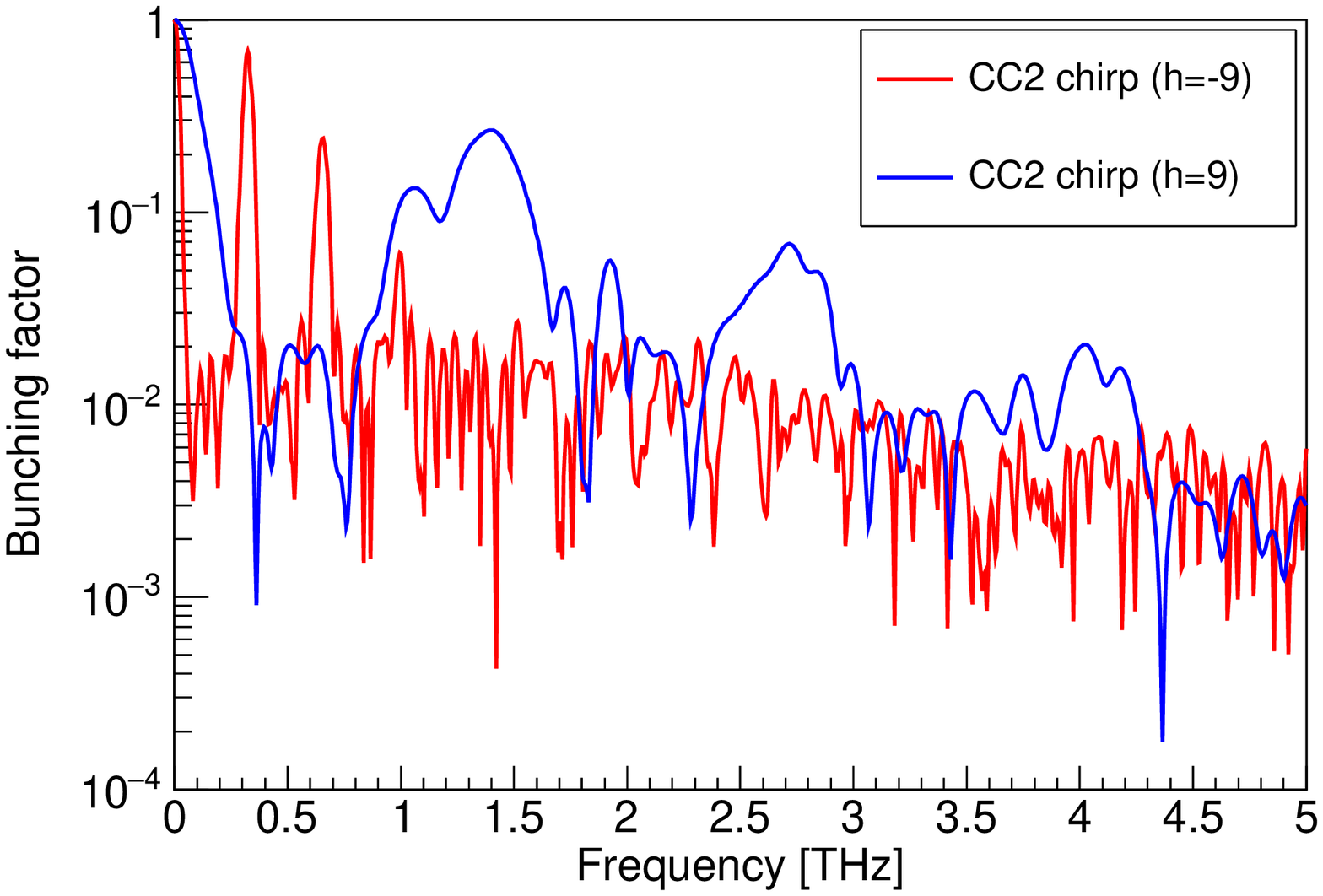}   
 \end{center}
\end{minipage}
\end{tabular}
\centering
      \includegraphics[width=6.0cm, height=5.0cm]{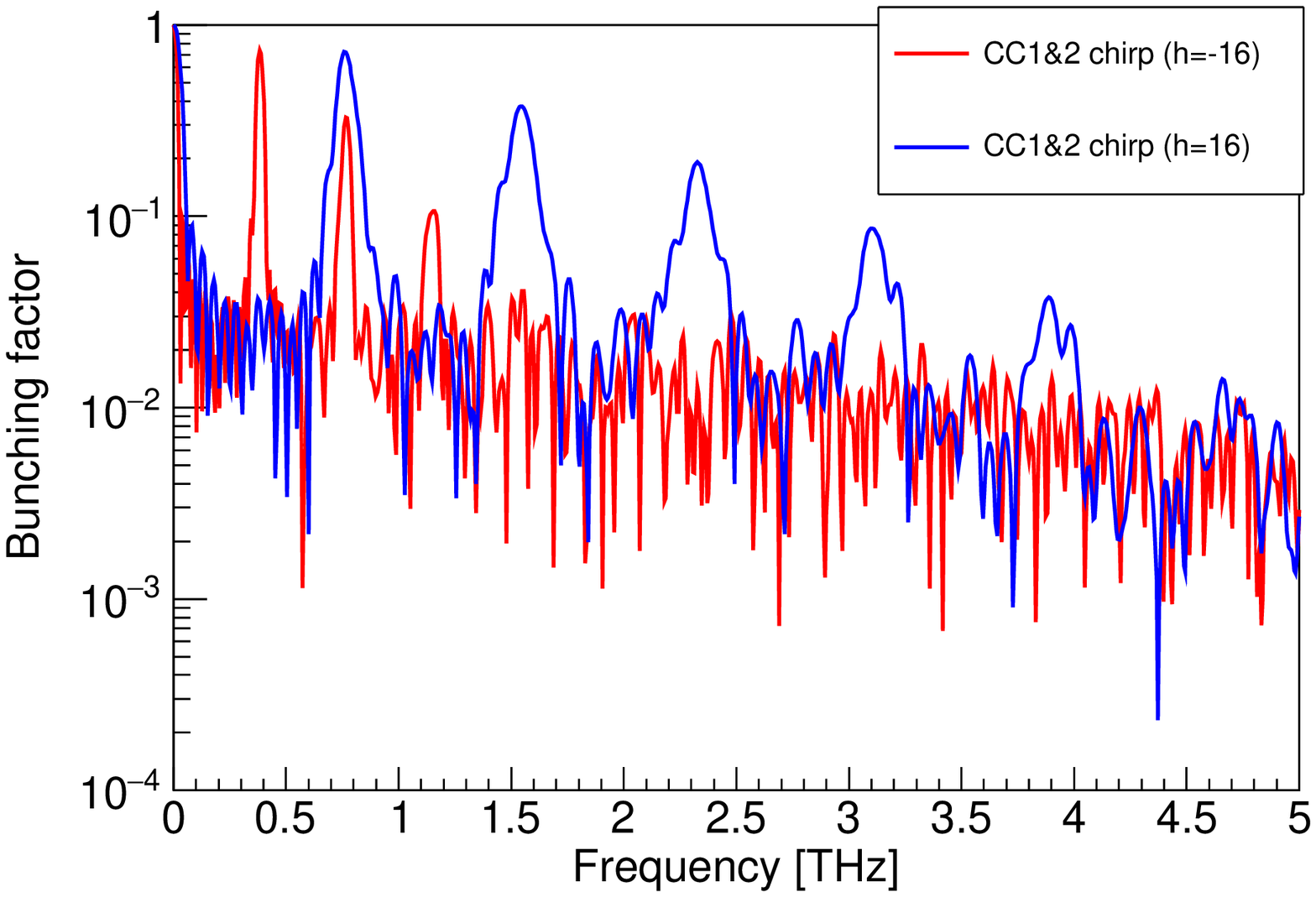}   
\caption{Bunching factors for CC1 chip ($h=\pm7\,\mathrm{m^{-1}}$), CC2 chirp ($h=\pm9\,\mathrm{m^{-1}}$), and CC1\&CC2 chips ($h=\pm16\,\mathrm{m^{-1}}$). Red lines are the lengthening mode, and blue lines are the over-compressed mode.}
\label{fig:5-2}
\end{center}
\end{figure}
We performed particle tracking with {\sc elegant} including effects of the slit-mask, magnet nonlinearities, 
longitudinal space charge effects, and coherent 
synchrotron radiation (CSR) in the chicane. Figure \ref{fig:5-1} shows the longitudinal distributions for 
$h=\pm7\,\mathrm m^{-1}$ (CC1 chirp), 
$h=\pm9\,\mathrm m^{-1}$ (CC2 chirp), and $h=\pm16\,\mathrm m^{-1}$ (CC1\&2 chirps) at X121. Table\,\ref{tb:1} shows 
the width, micro-bunch spacing, and fundamental frequency obtained both by particle tracking and analytical calculations with 
Eq.\,(\ref{eq:sigma_MB}, \ref{eq:3-2-4}),  and (\ref{eq:3-3-1}). The longitudinal distributions at $h=$-7, -9, and 
$\pm16\,\mathrm m^{-1}$ are separated clearly but not at $h=$7 and 9 $\mathrm m^{-1}$. Moreover, the spacing and 
width of micro-bunches at $h=$-7, -9, and $\pm16\,\mathrm m^{-1}$ obtained by particle tracking  agree with those 
computed with Eq.\,(\ref{eq:3-2-4}) and (\ref{eq:sigma_MB}). For the two cases of $h=$7 and 9 $\mathrm m^{-1}$  
(over-compressed modes), the overlap between micro-bunches are caused by the small separation $\lan \Delta z\ran $. Then,
the width of micro-bunches is difficult to estimate from particle tracking correctly due to the large overlap.

Figure \ref{fig:5-2} shows the bunching factor obtained from FFTs of the longitudinal distributions for different energy 
chirps: $h$=$\pm$7, $\pm$9, and $\pm$16 $\mathrm m^{-1}$. At $h=$-7, -9 ,and $\pm16$ $\mathrm m^{-1}$, narrow 
band frequency spectra are obtained due to well separated micro-bunches (see Fig.\,\ref{fig:5-1}). The 
fundamental frequencies at the first peak obtained by particle tracking including longitudinal space charge effects are 
consistent with the results from
Eq.\,\ref{eq:3-3-1}. On the other hand, at $h=$7 and 9 $\mathrm m^{-1}$ where the micro-bunches overlap, 
the two frequency spectra have broad peaks 
and there are differences in the fundamental frequencies between the simulations and the analytical results. 
 However, the spectra for positive chirps just above maximum compression are not narrow-band. These results show that 
using both cavities to create the chirp, i.e. $h=\pm 16$ m$^{-1}$, is most useful to create
high frequency narrow-band THz radiation. 
   \begin{table}[t]
   \centering
   \caption {Micro-bunch widths, spacing, and fundamental frequencies for each energy chirp. 
$\phi_{i, rf}=0^\circ$ implies on-crest RF phase. Values in parentheses show the results with longitudinal space charge (LSC),
and calculated analytically with Eq.\,\ref{eq:3-2-4}, \ref{eq:sigma_MB}, and \ref{eq:3-3-1}, respectively.
Initial bunch charge was 200 pC. }
   \label{tb:1}
 \begin{tabular}{ccccc} \hline
 RF phase & Energy chirp & Width & Spacing & Fundamental freq. \\ \hline 
   $\phi_{1, rf}$, $\phi_{2, rf}$ [deg.] & [$\rm{m^{-1}}$] &[mm] &[mm] & [THz] \\  \hline 
   (+35 , +35) & -16  & (0.09, 0.09) & (0.78, 0.72) & (0.39, 0.42) \\
   (0 , +35)     &   -9  & (0.12, 0.11) & (0.92, 0.87) & (0.33, 0.35) \\
   (+35 , 0)     &   -7  & (0.13, 0.12) & (1.01, 0.95) & (0.30, 0.32) \\
   (-35 , 0)      &    7  & (0.04, 0.02) & (0.13, 0.11) & (2.33, 2.73) \\
   (0 , -35)      &    9  & (0.06, 0.03) & (0.22, 0.19) & (1.37, 1.54) \\
   (-35 , -35)   &   16 & (0.04, 0.04) & (0.39, 0.34) & (0.76, 0.87) \\ \hline
   \end{tabular}
   \end{table}

Longitudinal space charge effects appear to have a negligible impact when the initial bunch charge is 200 pC.
For example, the changes in the fundamental frequency are about 1\% and there are no discernible differences in the
bunching factor shown in Fig.\,\ref{fig:5-2} with or without LSC. At a bunch charge of 1 nC, the higher harmonics beyond the
third get broadened and not as well defined as without the LSC inclusion. These higher harmonics will likely be beyond the
high frequency cutoff imposed by vacuum windows and not of practical relevance. 

\section{CTR and Wiggler Radiation Spectra}  \label{sec: radenergy}

In this section we examine and compare the spectra from two different radiation sources: transition radiation from an Al foil
and wiggler radiation. While both of these produce broad-band radiation, the bunching factor of the micro-bunched beam 
considered in the previous section results in narrow band radiation which is tuned by varying the chirp in the cavities CC1 and
CC2. This radiation is coherent at frequencies $ f \le c/\sg_z$ where $\sg_z$ is the micro-bunch rms width. 
In this range of  frequencies, the differential energy spectrum for  a bunch of $N$ electrons  is given in terms of the
single particle spectrum by
\beqr
\left(\fr{d^2 U}{d\Om d\om}\right)_N = \left(\fr{d^2 U}{d\Om d\om}\right)_1 [N + N(N-1) S(\om)],
\eeqr
where $U$ is the energy, $\Om$ is the solid angle, $\om$ the angular frequency and $S(\om)$ denotes the bunching factor.  

The transition radiation spectrum for a single electron moving through an infinite metallic foil is given by the well known
Ginzburg-Franck expression
\beqr
\left(\fr{d^2 U}{d\Om d\om}\right)_1 = \fr{e^2}{4 \pi^3 c \eps_0} \fr{\bt^2 \sin^2\theta}{(1 - \bt^2 \cos^2 \theta)^2},
\label{eq: dif2U_GF}
\eeqr
where $\bt$ is the relative velocity and $\theta$ is the horizontal angle of observation. This expression is independent of
the frequency. Modifications to this spectrum due to the finite size of the foil were derived in \cite{1-12}
for detection both in the near field and far field. At FAST, the detector will be about 1.5 m from the source placing it in the
far field at 1 THz. The radius of the foil is 1.25 cm which is comparable to the effective source size $\gm \lm$ at 1 THz.
We find that the foil size modifications to the spectrum are quite small, so we use the Ginzburg-Franck expressions in the
following calculations.  Integrating Eq.\,(\ref{eq: dif2U_GF}) over the solid angle, the $N$ particle ($N \gg 1$) differential energy spectrum
with respect to frequency is 
\beqr
\left(\fr{d U_{CTR}}{d\om}\right)_N \simeq N^2 \fr{e^2}{4 \pi^3 c \eps_0}\left[ \fr{1 + \bt^2}{\bt}\ln(\fr{1+\bt}{1-\bt}) - 2\right] S(\om)
\eeqr
Here $N$ is the number of particles in each micro-bunch. Using the bunching factors shown in Fig.\,\ref{fig:5-2}, the
energy density in $\mu$J/THz for two chirp settings using both CC1 and 2 are shown in Fig.\,\ref{fig: dU_CTR_Wiggler}. Here we
assumed an initial bunch charge of 1 nC and 5\% transmission, so that $N= 50$ pC at the foil. We discuss the possibility of 
choosing a mask with wider slit openings and higher transmission but lower bunching factor later in this section. 
The left plot in Fig.\,\ref{fig: dU_CTR_Wiggler} shows that the energy density at the first harmonic with either chirp setting is 
about 0.15 $\mu$J/THz. 
Three harmonics are visible up to $\sim3$ THz for both settings, but the
energy density falls more slowly with frequency for the compression setting ($h=+16$), as expected. 
The peaks in the spectrum are at (0.39, 0.75, 1.14) THz with bunch lengthening and at (0.75, 1.53, 2.14) THz with bunch
compression. 

Instead of CTR, it is conceivable to use a wiggler as a broad-band source of radiation which has the advantage of higher 
photon flux but requires more space in the beamline. Here we provide an estimate of the energy density expected from a
wiggler and compare it with the energy density from CTR. Using Eq.\,(3.19) in \cite{Kim_89} and integrating over the
horizontal angle, we find that the single particle spectrum in a bending magnet is
\beqr
\fr{dU_{bend}}{d\om} & = & \fr{\sqrt{3}}{2} \hbar \al_f \gm G_1(\om/\om_c)  \\
G_1(y) & = & y \int_y^{\infty} K_{5/3}(x) dx, \;\;\; \om_c = \fr{3}{2} \fr{\gm^3 c}{\rho}  \nonu
\eeqr
Here $\al_f$ is the fine structure constant, $K_{5/3}$ is a Bessel function, $\om_c$ is the critical frequency, and
$\rho$ is the bend radius of the magnet. A plot of the function $G_1$ can be seen in Fig.\,3.2 in \cite{Kim_89}. 
In the approximation that the radiation from a wiggler can be viewed as 
the radiation from a series of $N_P$ bending magnets (number of periods = $N_P/2$), the coherent differential energy
spectrum from $N$ electrons  ($N \gg 1$) going through a  wiggler, for frequencies $\om < 2\pi c/\sg_z$ can be approximated as
\beqr
\left(\fr{d U_{\rm{Wiggler}}}{d\om}\right)_N \simeq \fr{\sqrt{3}}{2} \hbar \al_f \gm N^2 N_P  G_1(\om/\om_c)  S(\om), 
\label{eq: dUwigg}
\eeqr
If $\lm_W$ is the length of the wiggler period, then the parameter defining the wiggler strength is $K = 0.934 \lm_W[{\rm cm}] B_0[{\rm T}]$ where
$B_0$ is the bending field. Typically $K > 2.5$ describes the transition from a multiple harmonics undulator radiation to the 
broadband wiggler radiation. 
For significant THz radiation, we require a low critical frequency $\om_c$ and a compact wiggler requires small values of 
$\lm_W, N_P$. Choosing for an example calculation, $B_0 = 0.2$ T yields the critical frequency $f_c = 40.5$ kHz at the 
FAST energy and with $\lm_W = 15$ cm, $K = 2.8$ and $N_P=10$ results in a wiggler length of 1.5 m. 
The right plot in Fig.\,\ref{fig: dU_CTR_Wiggler} shows the energy density spectrum using Eq.\,(\ref{eq: dUwigg}) and the
 bunching factor calculated above.
\bfig
\centering
\includegraphics[width=6.0cm, height=5.0cm]{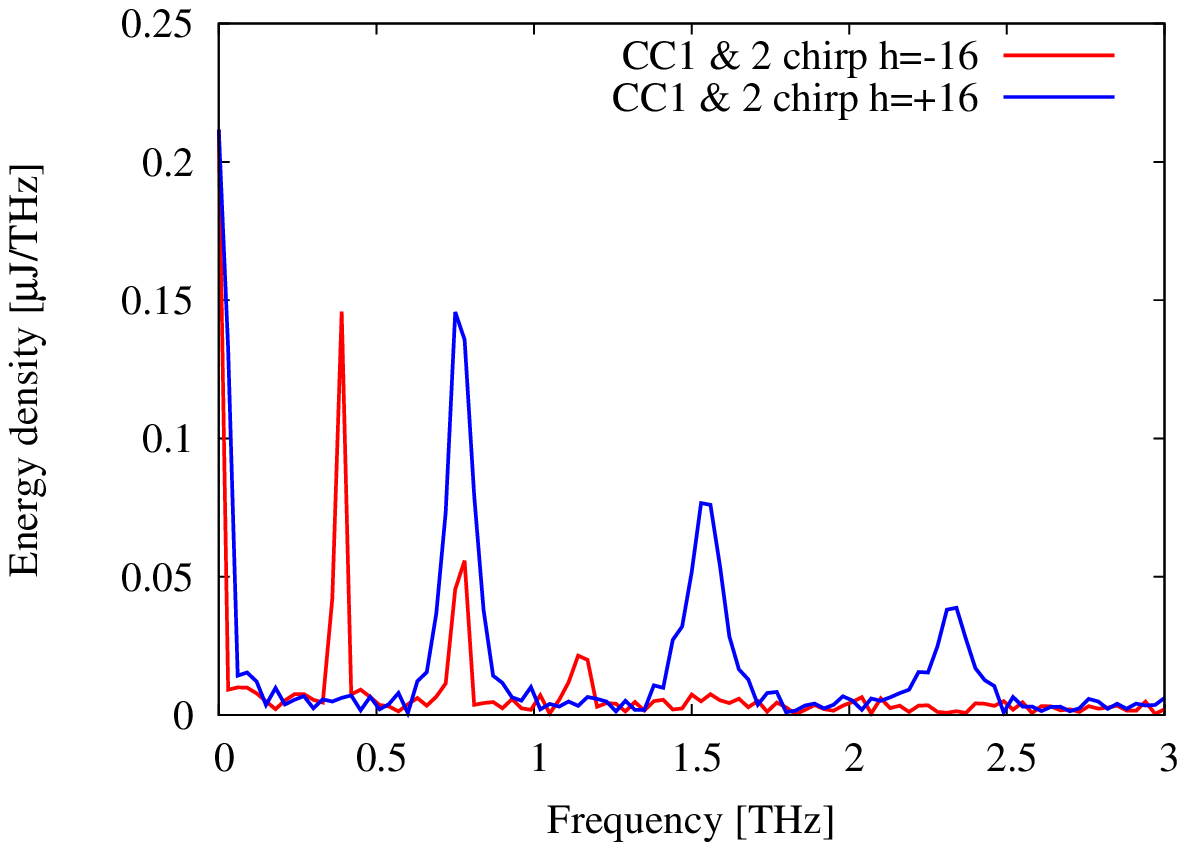}
\includegraphics[width=6.0cm, height=5.0cm]{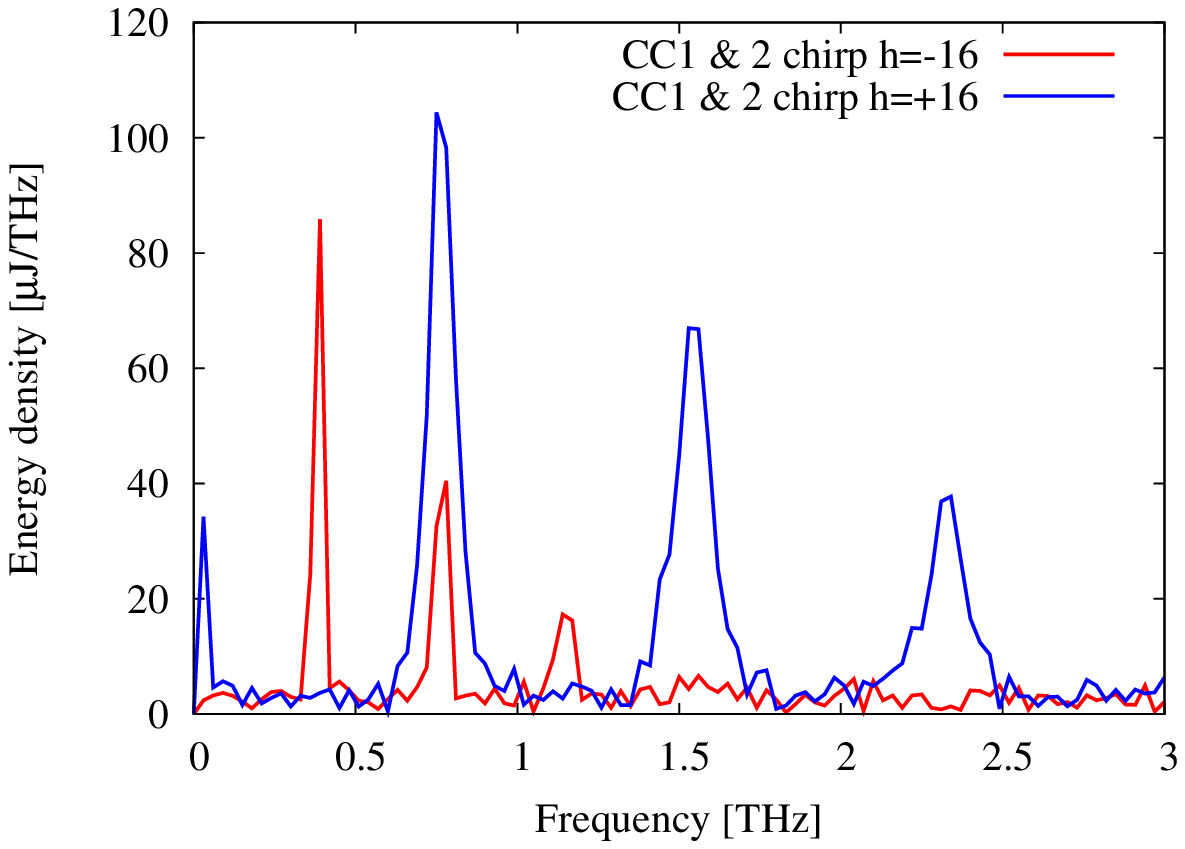}
\caption{Energy density spectrum from two radiation sources with two chirp settings using CC1 and CC2. 
Charge per bunch at the radiator is 50 pC. Left: From CTR, Right: From a wiggler, $K=2.8$, length = 1.5 m. }
\label{fig: dU_CTR_Wiggler}
\efig
The energy density with the wiggler at the first harmonic reaches (86, 104) $\mu$J/THz for $h=(-16, 16)$ respectively, 
nearly three orders of magnitude higher than from CTR. 
However, since the angular spread of wiggler radiation is larger
($\sim K/\gm$) than that of CTR ($\sim 1/\gm$), the energy deposited from a wiggler  can be expected to be about two orders of
magnitude larger. 

The slit width of the mask installed in the FAST beamline was chosen to be 50 $\mu$m spaced apart by 950 $\mu$m, 
primarily to  increase the separation between the micro-bunches and have a large bunching factor of order one. This also 
results in a low transmission
of 5\% to the radiator. Since the coherent radiation energy scales with the square of the
bunch charge and linearly with the bunching factor, it may be possible to increase the radiated energy from the above 
estimates by increasing the slit width for greater transmission which will also increase the overlap between the micro-bunches
and reduce the bunching factor. This optimization can be done for the next iteration of the experiment with a different choice of
slit mask. 

We note that the bunching factor could be increased by using a longitudinal space charge amplifier 
(LSCA) configuration \cite{Schneid_Yurkov_2010}. The LSCA scheme relies on the fact that the longitudinal space-charge 
impedance has a broad maximum approximately centered at the wavelength $\lm_{opt}= 2 \pi \sg_{\perp}/\gm$ thereby 
resulting in energy modulation at wave vector amplitude around $k_{opt} = \gm/\sg_{\perp}$. Consequently, as the electron 
bunch propagates over a length $L_d$ while maintaining a transverse beam size $\sg_{\perp}$, it will accumulate significant 
energy modulation at the wavelength $\lm_{opt}$. This modulation can then be transferred to a longitudinal-density modulation 
via a dispersive section with a properly selected longitudinal dispersion $r_{56}$.  This amplification process can 
be repeated over several stages and is characterized by a single-stage gain which, in the linear regime, takes the form 
$G(k) \simeq 4\pi (I_0/I_A)  L_d  |Z(k)|/(\gm Z_0 ) |r_{56} | {\cal C} k \exp\left [-({\cal C} k r_{56} \sg_{\dl} )^2/2\right]$, where 
${\cal C} \equiv (1 + h r_{56})^{-1}$ is the compression factor, $I_0$ and $I_A \simeq 17$ kA are 
the peak and Alfven currents respectively, $Z_0$ is the free-space impedance. For our set of parameters, and considering the 
case where we wish to amplify a wavelength $\lm_{opt} \simeq 300$~$\mu$m (1 THz) while selecting $\sg_{\perp}$ to be at 
the peak of the impedance so that  $|Z(k)|/Z_0 \sim 1$, we find that the single-pass gain to be approximately 
$G \sim 0.5 L_d$. Therefore, a propagation 
distance $L_d \sim 20$ m would be required to provide significant gain ($G \sim 10$). We do not consider this possibility here 
as it requires significant drift space with adequate optics to maintain the beam focused to the optimum spot size 
$\sg_{\perp}$  and 
would require the addition of a small chicane. Nevertheless, we should point out combining this LSCA technique with our slit 
approach could also enable the selection of a wider slit size thereby increasing the overall transmitted charge and the final 
THz-radiation signal. 

\section{Simulations of micro-bunched beam observation}
\label{obs}
\begin{figure}[t]
\begin{center}
  \begin{tabular}{c}
  \begin{minipage}{0.5\hsize}
\centering
\includegraphics[width=6.0cm, height=5.0cm]{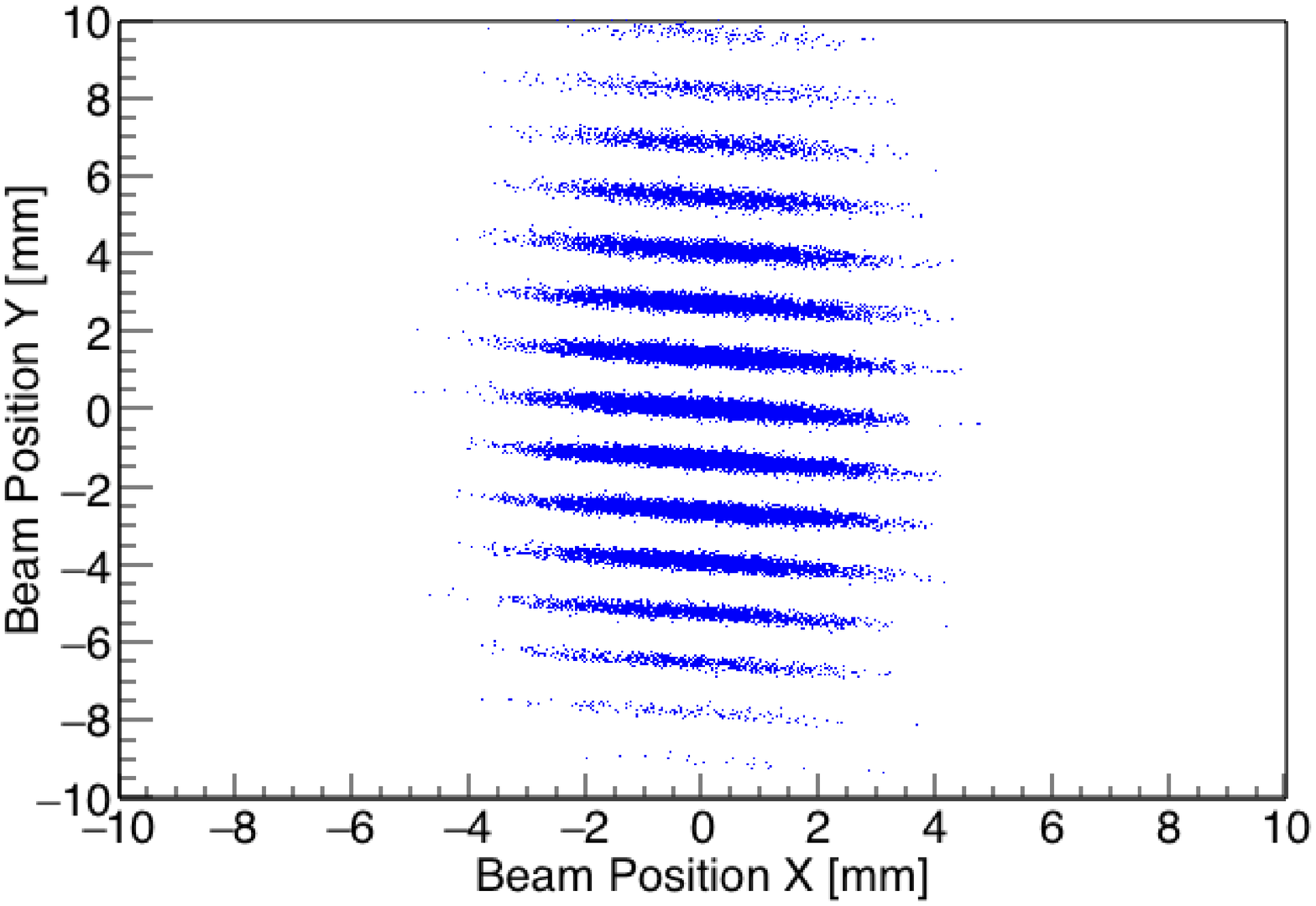}
\end{minipage}
  \begin{minipage}{0.5\hsize}
\centering
 \includegraphics[width=6.0cm, height=5.0cm]{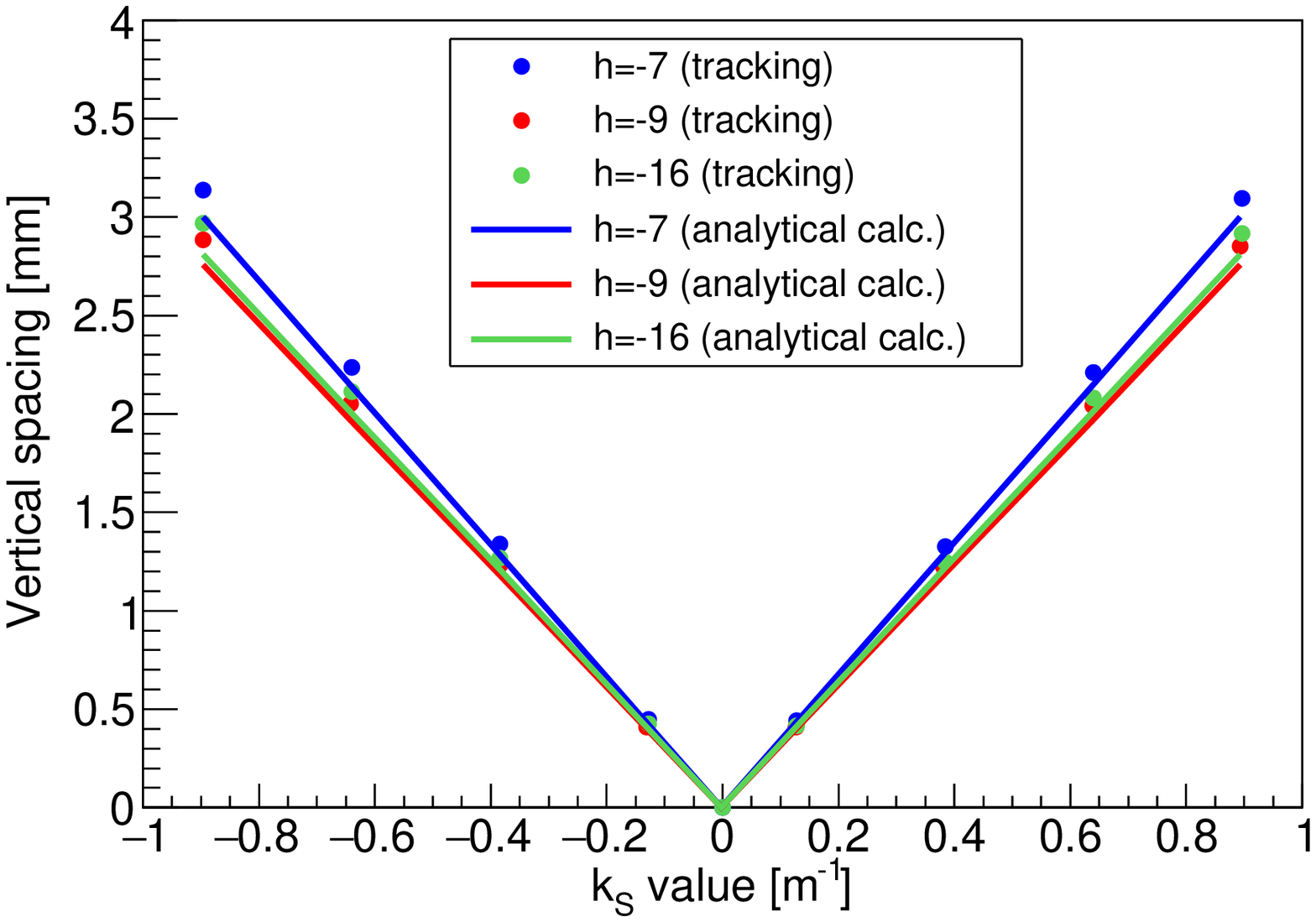}
\end{minipage}
\end{tabular}
\caption{Left: The transverse distribution at $h$=-7 ${\mathrm m^{-1}}$ at $X$121 when the skew quadrupole 
strength $k_S=  -0.39$ m$^{-1}$. 
Right: Vertical spacing as a function of $k_S$. Dots and lines represent the results obtained by particle tracking and analytical calculation, respectively.}
    \label{fig:6-1}
\end{center}
\end{figure}

As discussed in Section \ref{Micro}, a skew  quadrupole in the chicane generates vertical dispersion given by 
$\eta_y = -  k_S RM_{34}\eta_x$  where $\eta_x$
is the horizontal dispersion at the skew quadrupole. As a result,  the micro-bunches are vertically separated after the chicane. 
The left plot in Figure\,\ref{fig:6-1} shows the transverse distribution at X120 for $k_S$=-0.39 $\mathrm m^{-1}$ for $h$=-7 $\mathrm{m^{-1}}$ . The distribution is tilted to the right due to the beam coupling. The transverse distributions are  similar and the
vertical spacings are the same for other chirp values, as predicted by Eq.\,(\ref{eq:3-4-3}).  The right plot in Figure\,\ref{fig:6-1} 
shows the vertical spacing of the electron beam's dependence on the strength of the skew quadrupole from particle tracking and 
from Eq.\,(\ref{eq:3-4-3}). The vertical spacing computed with Eq.\,(\ref{eq:3-4-3}) is consistent with that from 
particle tracking. Also, the vertical spacing is proportional to the skew quadrupole strength as shown by this equation. 

\section{Conclusions}
\label{con}
In this paper, we have presented  theory and simulation results related to the THz radiation experiments planned at the FAST 
injector. We showed that narrow-band THz radiation with a frequency of over 1 THz can be 
generated from a micro-bunched beam using a slit-mask placed in the chicane. Particle tracking was done using {\sc elegant} and 
effects of magnet nonlinearities, CSR and LSC  were included. We showed that the emitted frequencies can be changed by 
varying the energy chirps (RF phases) in the accelerating cavities. This scheme for generating narrow band tunable THz
radiation  is relatively simple and does not require  either laser modulation of the electron bunch  or variable gap undulators. 

In general, lengthening the bunch with negative chirp settings results in larger separations of the micro-bunches after the
chicane and spectrum peaks with narrower widths. At positive chirp settings close to the maximum compression  
($h = 5.6$ m$^{-1}$), there is large  overlap between the micro-bunches resulting in a broad-band spectrum. 
However,  at sufficiently
large positive chirp e.g. $h=16$ m$^{-1}$, the micro-bunch widths $\sg_{z,MB}$ and separations $\lan \Delta z\ran$ are 
smaller than for $h=-16$ m$^{-1}$ but nevertheless obey $\sg_{z,MB}\ll \lan \Delta z \ran$, so that
the spectrum peaks are well separated. 
Chirping in both cavities is required for either of $h = \pm 16$ m$^{-1}$. The advantage with the large negative chirp is the 
spectral peaks are narrower,  the disadvantage is that the spectrum does not reach the higher frequencies 
obtained by compressing the bunch with the large positive chirp. We note that the negative chirp case is somewhat more 
operationally efficient since a bunch exits the rf gun with a negative chirp (i.e. higher energy particles are at the head) due to 
longitudinal space charge forces within the gun cavity.

A CTR foil will be used in the FAST beamline to generate the THz radiation and we calculated the expected radiation 
spectra for $h= \pm 16$ m$^{-1}$. Assuming a charge of 50 pC reaches the radiator, the energy density at the first 
harmonic for either chirp is $\sim 0.15 \; \mu$J/THz. Using a relatively compact wiggler, we found that the energy density
would be about three orders of magnitude higher. For both radiation sources, the power in the higher harmonics is 
significantly higher for bunch compression with $h= +16$ m$^{-1}$. 

In the initial stage we plan to use a skew quadrupole downstream of the slit-mask in the chicane to observe  micro-bunching in the vertical 
plane.  The vertical spacing at  a monitor downstream of the chicane is shown to be proportional to the skew quadrupole 
strength. We found that it is necessary to focus the beam vertically at the monitor to obtain clear vertical separations of
the slit images but too strong focusing results in chromatic distortions of the image. However chromatic effects are
quite weak when the beam is focused to spot sizes of $\sim1$ mm (resulting in a sufficiently large high frequency cutoff 
$\sim$3.3 THz) at the Al target for THz production. 

\section{Acknowledgments}
The first author would like to express his gratitude to 
SOKENDAI (The Graduate University of Advanced Studies) and KEK for  ``INTERNSHIP PROGRAM IN 2017" and to Fermilab for supporting his studies. We also thank the referee for helpful suggestions.  This manuscript has been authored by Fermi Research Alliance, LLC under Contract No. DE-AC02-07CH11359 with the U.S. Department of Energy, Office of Science, Office of High Energy Physics.

% \section* {References}

\end{document}